\documentclass{osa-article}

\journal{osac}

\articletype{Research Article}

\usepackage{lineno}

\begin{document}

\title{Flatfield Ultrafast Imaging with Single-Shot Non-Synchronous Array Photography}

\author{Matthew Sheinman,\authormark{1,*} Shyamsunder Erramilli,\authormark{1,2} Lawrence Ziegler,\authormark{2,3} Mi K. Hong,\authormark{1} and Jerome Mertz\authormark{3,4}}

\address{\authormark{1}Department of Physics, Boston University, 590 Commonwealth Avenue, Boston, Massachusetts 02215, USA}

\address{\authormark{2}Photonics Center, Boston University, 8 St. Mary’s St., Boston, Massachusetts 02215, USA}

\address{\authormark{3}Department of Chemistry, Boston University, 590 Commonwealth Avenue, Boston, Massachusetts 02215, USA}

\address{\authormark{4}Department of Biomedical Engineering, Boston University, 44 Cummington Mall, Boston, Massachusetts 02215, USA}

\email{\authormark{*}sheinman@bu.edu} 



\begin{abstract}
We present a method for acquiring a sequence of time-resolved images in a single shot, called Single-Shot Non-Synchronous Array Photography (SNAP). In SNAP, a pulsed laser beam is split by a diffractive optical element into an array of angled beamlets whose illumination fronts remain perpendicular to the optical axis. Different time delays are imparted to each beamlet by an echelon, enabling them to probe a scene in rapid succession. The beamlets are then imaged onto different regions of a camera by a lenslet array. Because the illumination fronts remain flat (head-on) independently of beamlet angle, the temporal resolution of SNAP is fundamentally limited only by the laser pulse duration, akin to a "global shutter" in conventional imaging. We demonstrate SNAP by capturing the evolution of a laser induced plasma filament for 20 frames at an average rate of 4.2 Tfps.
\end{abstract}

\section{Introduction}

Ultrafast imaging has the potential to reveal fundamental mechanisms in chemistry, physics, and biology\cite{gorkhover2016femtosecond,imada1998metal,tuchin2007methods}. Often, ultrafast dynamics are measured by repeated acquisitions. \cite{papazoglou2008line,wang2015time,hu2015comparison,vsiaulys2015situ}. However, this approach is not feasible for events that are difficult or impossible to reproduce, such as optical rogue waves\cite{solli2007optical}, irreversible phase transitions\cite{poulin2006irreversible}, or laser-plasma interactions\cite{li2016single}, thus motivating the development of techniques that are single shot.

Compressed imaging techniques have been successful at capturing ultrafast scenes in a single shot. Compressed Ultrafast Photography (CUP) has the advantage of being completely passive, that is, it can capture light emitted from an ultrafast scene without the need of a structured probe beam \cite{gao2014single}. However, the compressed nature of CUP imposes a sparsity requirement for image reconstruction. In contrast, a variety of approaches allow the direct probing of ultrafast events in a single shot without the need for reconstruction\cite{Liang:18,faccio2018trillion,mikami2016ultrafast}. 

One approach involves spectral encoding, which associates portions of an ultrafast pulse's spectrum to different times.  This approach enables a number of temporally resolved images to be captured in a single acquisition with high spatial resolution. A representative such technique is sequentially timed all-optical mapping photography (STAMP), or its successor, SF-STAMP, which can capture 25 frames in a single shot \cite{nakagawa2014sequentially,suzuki2017single}. However, because each image (time point) is probed only by a portion of the original spectrum, temporal resolution degrades as the number of images increases.

Another approach associates a unique spatial frequency to each time. This disassociates temporal resolution from the number of frames, since each image is acquired with the full pulse spectrum. In FRAME, a scene is probed by a train of pulses with discrete spatial frequencies \cite{ehn2017frame}. Other approaches utilize digital holography \cite{moon2020single,huang2019single}. In particular, Time and Spatial-Frequency Multiplexing (TSFM) has been used to capture up to 14 images in a single shot, and could potentially capture more images relatively easily. TSFM uses a diffractive optical element (DOE) to produce an array of reference beams for off-axis holography, where each reference beam is assigned a unique time delay by an echelon. However, because spatial frequency techniques involve multiplexing, SNR becomes degraded as the number of images increases.

Yet another approach involves associating incidence angle to time, which has the advantage of avoiding loss of temporal resolution while also not requiring multiplexing. Angular encoding via pairs of two-by-two mirror arrays has been demonstrated\cite{Yeola:18}, though this technique is difficult to generalize to larger frame numbers.

Here, we demonstrate that angular encoding by SNAP can easily capture a large number of frames \cite{sheinman2021trillion}. As in TSFM, we use a DOE and an echelon to produce an array of time delayed beamlets. However, instead of using these beamlets as reference beams for holography, we probe the scene directly. Because the DOE preserves flat illumination fronts, each beamlet is incident on the entire scene simultaneously\cite{huang2019single}. We then project the images from each beamlet onto different regions of a camera, where they are recorded in a single shot \cite{guo2019fourier}. As a demonstration, we captured 20 shadowgraph images of the formation of a laser induced plasma filament at an average framerate of 4.2 Tfps.  

We note that while finalizing our manuscript for submission, we became aware of very recent work by Zhu et al. based on a similar concept \cite{zhu2021single}, though without the benefit of flatfield (head-on) illumination provided by a DOE.

\section{Principle of SNAP}

\begin{figure}[h!]
\centering\includegraphics[width=13cm]{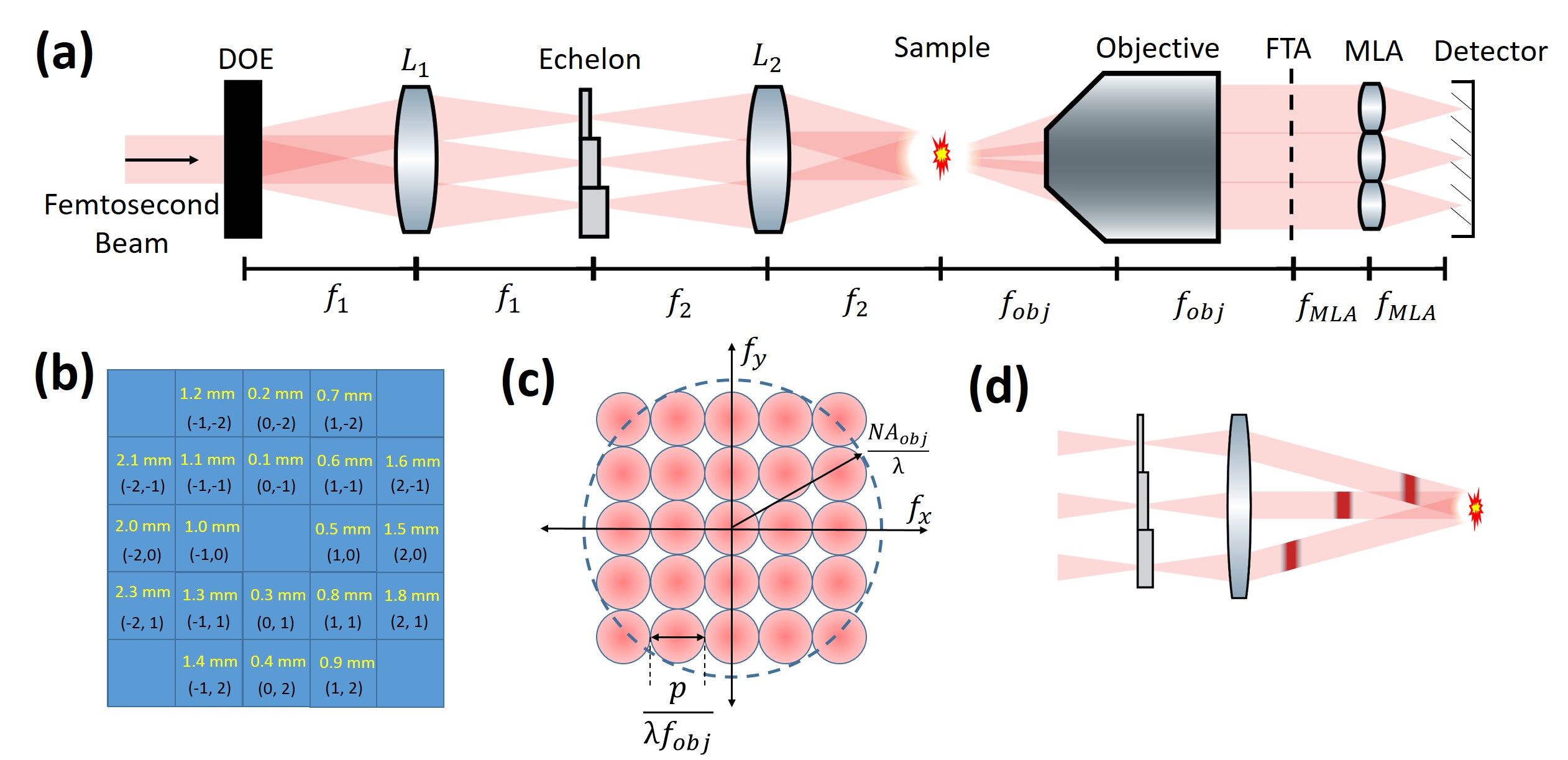}
\caption{Working principle of SNAP. (\textbf{a}) Schematic of setup. (\textbf{b}) The echelon as viewed head on. The black indices (i,j) identify individual sections; the yellow text denotes their thickness. (\textbf{c}) Fourier-plane array of beamlets (red circles). Dashed circle represents the objective frequency support. To maximize spatial resolution, the FPA should nearly fill the objective. (\textbf{d}) Probe pulses generated by the DOE maintain flat illumination fronts perpendicular to the optical axis. This prevents any "rolling shutter" effect, even at the large incidence angles required to maximize spatial resolution.}
\end{figure}

The layout of our system is illustrated in Fig. 1a. SNAP resolves time delays by probing a scene with an array of angles and then recording each of these angles individually. This can be understood through well known principals in optics\cite{mertz2019introduction}. A femtosecond pulse transmitted through a DOE produces an array of pulsed beamlets $a_{i,j}(t)e^{i\vec{k}_{i,j}\cdot\vec{\rho}}$. Here, $(i,j)$ are indices denoting diffraction order, and $a_{i,j}$ are amplitudes for spatially homogeneous ultrashort fields, but with intensities varying due to the efficiency of the DOE. The transverse vector $\vec{k}_{i,j}$ corresponds to the DOE diffraction angle $\theta_{i,j}$. 

As shown in Fig. 1a, the DOE is imaged onto a scene by a 4f relay formed by lenses $L_{1}$ and $L_{2}$. An echelon is placed at the Fourier plane of this relay (fig. 1b)\cite{poulin2006irreversible}. Each beamlet is focused through a unique section of this echelon, imparting the unique time delay

\begin{equation}
    \tau_{i,j} = \frac{d_{i,j}}{c}(n_{g}-1)
\end{equation}
where $d_{i,j}$ is the section thickness, $n_{g}$ is the group velocity of the echelon material, and $c$ is the speed of light in vacuum.

Immediately after the scene of the ultrafast event being imaged, the train of beamlets is given by

\begin{equation}
E_{sample}(\vec{\rho},t) = T(\vec{\rho},t)\sum_{i,j}a_{i,j}(t-\tau_{i,j})e^{i\vec{k}_{i,j}\cdot\vec{\rho}}
\end{equation}
where $T(\vec{\rho},t)$ is the complex amplitude transmittance of the ultrafast scene.

From the Fourier shift theorem the field at the objective back focal plane is then given by

\begin{equation}
E_{FPA}(\rho,t)=\sum_{i,j}a_{i,j}(t-\tau_{i,j})\tilde{T}(\frac{\rho}{\lambda f_{obj}}-\frac{k_{i,j}}{2\pi},t)
\end{equation}
\\That is, our array of angles has been transformed into an array of displacements in the objective's focal plane. The pitch $p_{FP}$ of the beamlet array in this place is

\begin{equation}
    p_{FP} = \frac{\lambda f_{obj}}{2 \pi}\Delta{k}=f_{obj}\frac{f_{1}}{f_{2}}\theta_{0}
\end{equation}
where $\Delta{k}$ is the magnitude of the difference between adjacent wave vectors and we have made use of the small-angle approximation.

To image each beamlet separately, we place a microlens array (MLA) in a 4f configuration with our objective. If the pitch of the MLA is chosen to closely match $p_{FP}$, the field at the camera plane becomes 


\begin{equation}
E_{camera}(\rho,t)=\sum_{i,j}a_{i,j}(t-\tau_{i,j})T(\frac{f_{obj}}{f_{MLA}}\rho - \frac{\lambda f_{obj}}{2\pi}k_{i,j},t)
\end{equation}

Finally, because our camera exposure time is orders of magnitude longer than our probe pulse duration, our signal is given by the time integral

\begin{equation}
I_{camera}(\rho)=\int_{-\infty}^{\infty}\sum_{i,j}|a_{i,j}(t-\tau_{i,j})|^{2}|T(\frac{f_{obj}}{f_{MLA}}\rho-\frac{\lambda f_{obj}}{2 \pi}k_{i,j},t)|^{2}dt
\end{equation}

Here we have assumed that our beamlets are well separated on the camera sensor, allowing us to omit cross-terms. Equation 6 describes an array of images of the ultrafast scene intensity transmittance at different times. The temporal resolution associated with each sub-image is simply the pulse duration $\Delta t_{i,j}$ of each corresponding probe beamlet, which differ slightly owing to dispersion from the echelon and the lenses $L_{1}$ and $L_{2}$.

Equations 5 and 6 are approximations that do not account for the finite aperture of each microlens. A more rigorous version of Eq. 5 includes a frequency cutoff defined by the MLA pitch $p$

\begin{equation}
    P(\xi) = 
    \begin{cases}
    1,&|\xi|<p/2
    \\0,&|\xi|>p/2
    \end{cases}
\end{equation}

Applying equation 7 to equation 3, we can infer the system's spatial resolution from the maximum spatial frequency $f_{max}$ associated with each sub-image
\begin{equation}
    f_{max} = \frac{p}{2\lambda f_{obj}}
\end{equation}

In effect, the objective's frequency support is divided between each sub-image (Fig. 1c). This introduces a trade-off between spatial resolution and number of sub-images, as observed in earlier angle encoding techniques \cite{Yeola:18}. To maximize spatial resolution then, the maximum probe beamlet tilt should approach the objective numerical aperture.

The advantage of using a DOE to impart angular tilts to the beamlets can now be understood (Fig 1c). The illumination fronts of the beamlets remain perpendicular to the optical axis, regardless of tilt angle, meaning that each beamlet is incident on the ultrafast scene head on. Without this advantage, the system would suffer from a "rolling shutter" effect which would become more pronounced for higher tilt angles. Indeed, with our existing setup, the most tilted probe beamlets generated without a DOE would require a full on the order of 1 ps to sweep through the FOV.

We note that Eq. 8 fails to take into account the vignetting of the FPA by our MLA. Vignetting can cause a loss of spatial resolution and temporal cross-talk between sub-images. We expect these effects to become more pronounced farther from the center of our FOV. In principle, such effects can be mitigated by departing from the 4f geometry and placing the MLA directly in the objective back focal plane, at the expense of increased aberrations. In our current setup, we elected to minimize aberrations and adopt the 4f geometry described above.

\section{Methods}

\subsection{Plasma Sample and Shadowgraphy}

\begin{figure}[h!]
\centering\includegraphics[width=14cm]{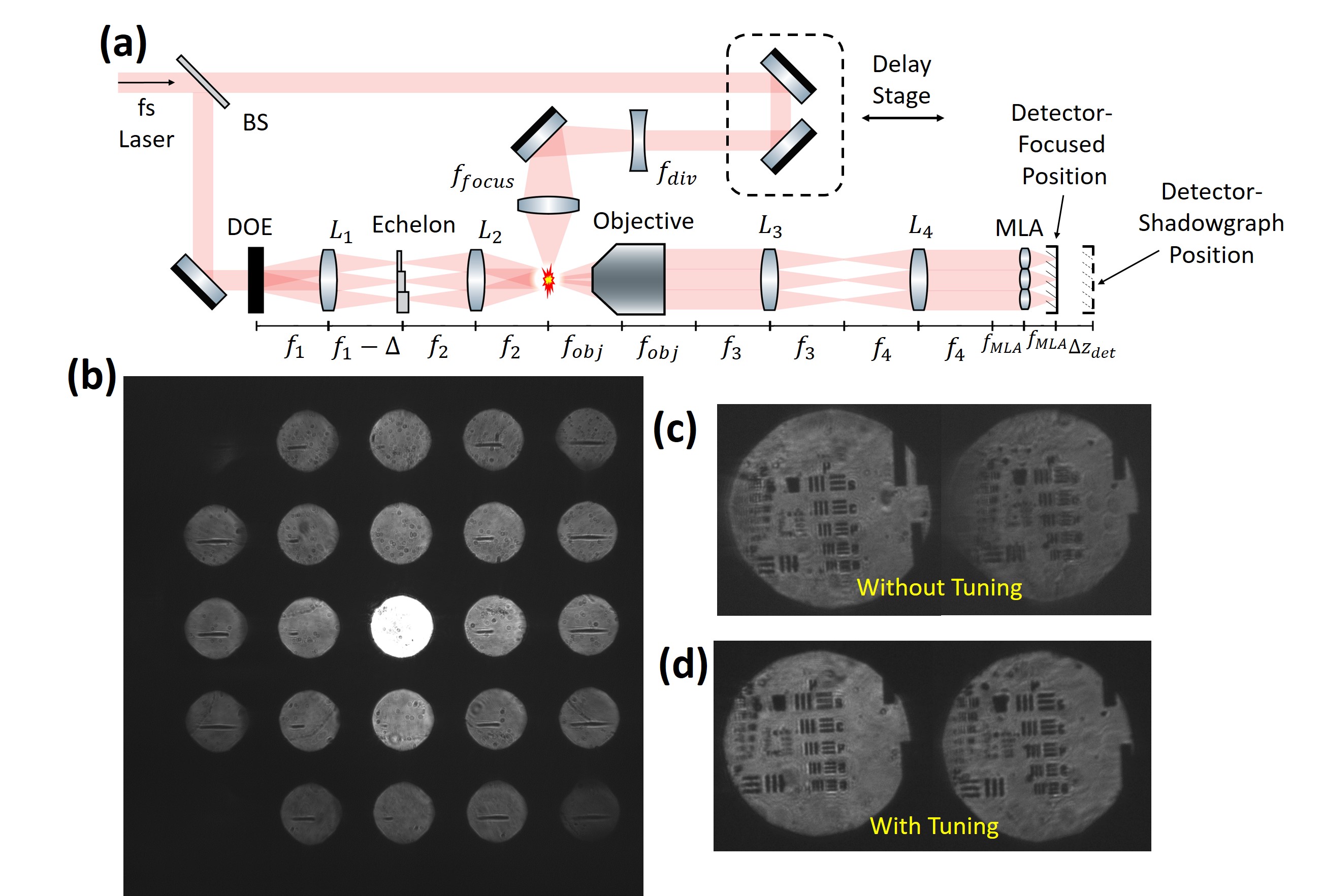}
\caption{Imaging with SNAP. (\textbf{a}) A more detailed schematic. The camera can be focused on the sample or displaced to produce a shadowgram. (\textbf{b}) Simultaneously acquired raw images of the plasma shadowgram at different time points. (\textbf{c}) Comparison of the resolution in sub-images (-1,-1) and (2,-1) when $\Delta$ is near 0. (\textbf{d}) Same sub-images when $\Delta$ is about 15 mm.}
\end{figure}
To demonstrate SNAP, we image the formation of a laser induced plasma in air. A diagram of the system's pump arm for inducing the plasma formation is shown in Fig. 2a. To better highlight SNAP's ability to image features in both space and time, the pump arm was designed to produce a plasma that exhibited spatial structure. Specifically, high energy pulses focused with a NA greater than 0.3 are known to produce plasma filaments with two distinct foci \cite{yu2016ultrafast}. Accordingly, a diverging lens is used to expand the pump beam and fill the focusing lens, thus increasing NA. In addition, the pump arm includes a delay stage to allow temporal calibration, as detailed below. Our pulse has a peak wavelength of 800 nm with a measured pulse duration of 50 fs.

To enhance the signal contrast produced by the plasma filament, we perform shadowgraph imaging by intentionally translating our camera approximately 3.5 mm beyond the nominal image plane (Fig. 2a). In this way, we image the diffraction pattern (shadowgram) caused by the filament rather than the filament itself.

To maximize SNR in as many sub-images as possible, we overexpose and discard the (0,0) image. Further discarding the four sub-images in the corners, we thus capture a total of 20 images, corresponding to 20 time points.

\subsection{Fourier Plane Tuning}

A more complete schematic of the system's imaging arm is shown in Fig. 2a. Additional relay lenses are placed between the objective and MLA. This serves the practical purpose of imaging the objective's focal plane, which would otherwise be inaccessible, onto the MLA's focal plane. In addition, this relay can be used to tune the value of $p_{FP}$ in front of the MLA.

We use a DOE with a diffraction angle of $\theta_{0}=9.3^{\circ}$ at 808 nm (HOLOEYE DOE 804). Our lenses have focal lengths $f_{1}$ = 50 mm, $f_{2} =$ 100 mm, $f_{3} =$ 100 mm, and $f_{4} = $ 70 mm. We use a $10\times$ objective and a MLA with pitch $p = 1$ mm, focal length $f_{MLA} = 4.8$ mm (Advanced Microoptic Systems APO-Q-P1000-R2.2). The displacement of the lens $L_{1}$ from its nominal 4f configuration by an amount $\Delta$ is explained below.

Equation 5 imposes a requirement on the relation between DOE angle $\theta$, MLA pitch $p$, and the optics chosen for the system. The above relay introduces a new degree of freedom to this relation:

\begin{equation}
    p_{FP} = f_{obj}\frac{f_{1}f_{4}}{f_{2}f_{3}}\theta
\end{equation}

In our case $p_{FP} = $ 1.02 mm, which is slightly larger than our MLA pitch, and may be responsible for irregularities in the resolution between sub-images (Fig. 2b). To further tune our system, we displaced the lens $L_{2}$ and DOE by an distance $\Delta$. With this displacement, a high degree of resolution uniformity can be achieved (Fig 2c). We speculate that this improvement is due to more refined tuning of the probe beam angles, and thus the pitch $p_{FP}$.

We note that, in addition to helping tune the pitch $p_{FP}$, displacing $L_{2}$ causes the probe beams to become divergent. For this reason, we expect the plasma shadowgram to expand more rapidly than the plasma itself.

\subsection{Temporal Calibration}

\begin{figure}[h!]
\centering\includegraphics[width=14cm]{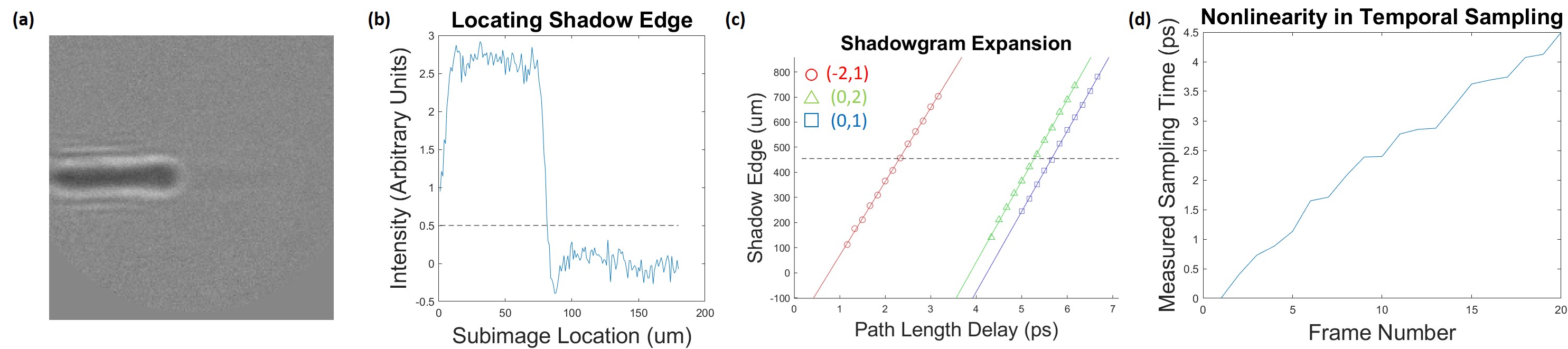}
\caption{Temporal calibration of SNAP. (\textbf{a}) A representative sub-image after background subtraction and normalization where the plasma has only partially formed. (\textbf{b}) A line profile (averaged over 8 rows) is used to identify the plasma leading edge. (\textbf{c}) Three of the 20 fits made for each sub-image, with the dashed line representing a reference point near the filament center. (\textbf{d}) Time stamps obtained from the temporal displacement of each of the 20 fits.}
\end{figure}

Our echelon is designed to provide an increase in glass thickness of 100 $\mu m$ per step and a group index $n_{g} = 1.53$. With these parameters, Eq. 1 predicts a time step of about 177 fs and frame rate of 5.6 Tfps. However, in the actual experiment, deviations from these values are observed, which we attribute to spherical abberations in the imaging relay formed by $L_{1}$ and $L_{2}$.

To correct for these deviations, we calibrate our system by measuring the relative arrival times of our probe beamlets. We do this by creating a conventional, many-acquisition ultrafast video for each sub-image. First, we acquire a background image by positioning our delay stage such that all beamlets arrive at the sample plane before the plasma begins to form. We then translate our delay stage in 25 $\mu m$ steps, each causing the plasma to form 167 fs later, and acquire the corresponding images. An example of one such image (background subtracted) is shown in Fig. 3a. The pump energy is about 2 mJ per pulse.

To extract time delays from this data, we identify the leading edges of the shadowgrams based on a threshold (Fig. 3b). The position of the leading edges is then plotted against the probe beamlet arrival times. Example linear fits are shown in Fig. 3c. The intersections of these fits with a location in the center of the FOV then provide a measure of the probe beam relative arrival times. These are plotted in Fig. 3d, enabling us to correct for the deviations from a linear trend.

For this method of temporal calibration to be effective, a common spatial axis must be found for each subimage. To achieve this, we place an obstruction in the plane of the plasma and image its shadowgram. We then perform normalized cross correlations between each of these shadowgrams, using subimage (0,1) as a reference image. With this choice of reference, the mean peak cross correlation value is .95.

\section{Results}

\begin{figure}[h!]
\centering\includegraphics[width=13cm]{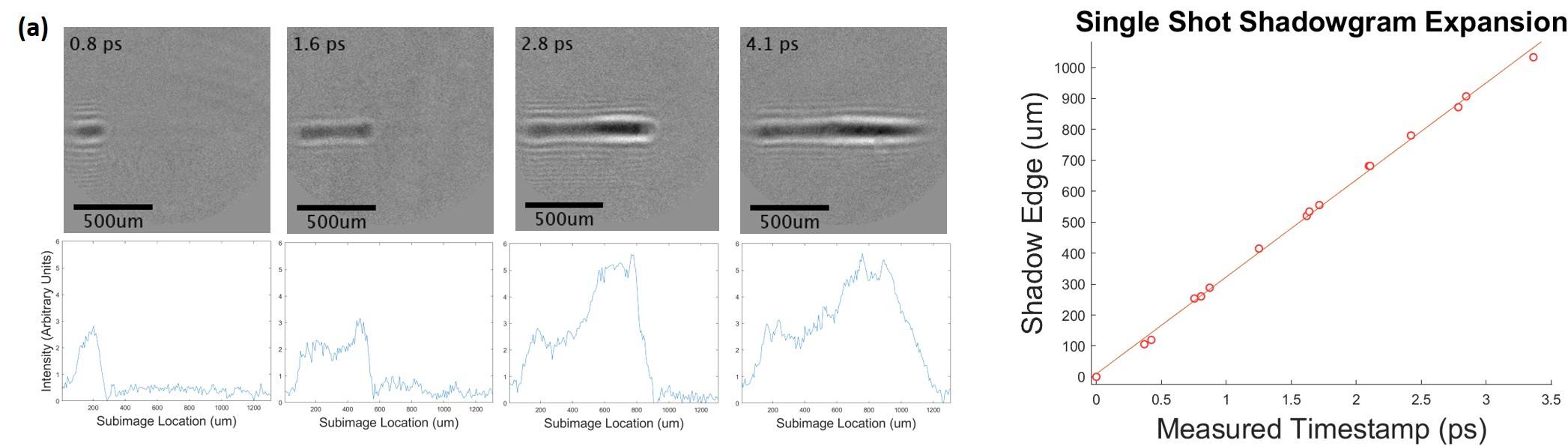}
\caption{Evolution of a plasma filament. (\textbf{a}) Four frames taken from our 20 sub-images, with associated line-scans (averaged over 8 rows). The sequential formation of two foci is evident. (\textbf{b}) A plot of the shadowgram leading edge progression based on calibrated time stamps.}
\end{figure}


A pump energy of about 1 mJ per pulse was found to produce laser plasmas with two clearly distinguishable foci, one leading and the other trailing. Example line scans from four of our 20 frames are shown in Fig. 4a. At early times when the plasma just begins to form and is highly localized, a sharp drop behind the plasma's leading edge is observed. As the plasma continues to form, a second focus becomes apparent. Finally, at long times, the sharp drop becomes a gradual roll-off at the edge of the second focus.

The single-shot data is fit to a line and compared with the data from the many-acquisition temporal calibration  (Fig. 4b). The single-shot expansion rate is observed to be 310$\pm$20 um/ps. The largest sources of uncertainty in our estimate are the shadow decay at the edge of our linescans and the uncertainty in spatial calibration by cross correlation.

To compare with conventional many-acquisition imaging, we take the average of the expansion rates measured during temporal calibration for the 8 inner subimages, where abberations should be at a minimum. We find a value of 318$\pm$5 um/ps, consistent with the single-shot value. Discrepancies with the known value of the speed of light are not examined, but may be due to the previously mentioned divergence in the probe beam.

\section{Conclusion}
In this work, we demonstrated that SNAP is a viable single-shot alternative to more conventional multi-shot ultrafast imaging techniques. There is a fundamental tradeoff in SNAP between spatial resolution and number of temporal samples. However, the temporal resolution of SNAP is limited only by the duration of the probe pulse. Importantly, the use of a DOE to generate the beamlet array means that the entire FOV of the ultrafast scene is probed simultaneously, analogous to a "global shutter".

We imaged the evolution of a laser induced plasma filament both with SNAP and with conventional many-acquisition imaging, and verified that the expansion rates of the resultant plasma shadowgrams were consistent. Moreover, SNAP was able to resolve a double focus structure in the filament, illustrating its effectiveness at capturing spatiotemporal data.

SNAP uses readily available optical components that are easy to assemble, and its generalization to a larger number of temporal samples is straightforward. As such, SNAP can be useful for a wide variety of  ultrafast imaging experiments where a high temporal resolution is required.

\section*{Funding}
National Institute of Health R21GM128020.


\bibliography{sample}






\end{document}